\documentclass[twocolumn,prl,aps,superscriptaddress,showpacs]{revtex4}

\usepackage{graphicx}
\usepackage{dcolumn}
\usepackage{amsmath}
\newlength{\figwidth}
\preprint{DRAFT}

\begin{document}


\title{Observation of a 500meV Collective Mode in
La$_{2-x}$Sr$_x$CuO$_4$ and Nd$_2$CuO$_4$}

\author{J. P. Hill}
\affiliation{Department of Condensed Matter Physics and Materials
Science, Brookhaven National Laboratory, Upton, New York 11973}
\affiliation{Department of Physics, Clarendon Laboratory, University
of Oxford, OX1 3PU, United Kingdom}
\author{G. Blumberg}
\affiliation{Bell Laboratories, Alcatel-Lucent, Murray Hill, New
Jersey 07974}
\author{Young-June Kim}
\author{D. Ellis}
\author{S. Wakimoto}
\author{R. J. Birgeneau}
\affiliation{Department of Physics, University of Toronto, Toronto,
Ontario M5S 1A7, Canada}
\author{Seiki Komiya}
\author{Yoichi Ando}
\affiliation{Central Research Institute of Electric Power Industry,
Komae, Tokyo, 201-8511, Japan}
\author{B. Liang}
\author{R.L. Greene}
\affiliation{Center for Superconductivity Research, Dept. of
Physics, U. of Maryland, College Park, Md 20742, USA}
\author{D. Casa}
\author{T. Gog}
\affiliation{CMC-XOR, Advanced Photon Source, Argonne National
Laboratory, Argonne, Illinois 60439}

\date{\today}

\begin{abstract}

Utilizing resonant inelastic x-ray scattering, we report a
previously unobserved mode in the excitation spectrum of
La$_{2-x}$Sr$_x$CuO$_4$ at 500 meV. The mode is peaked around the
($\pi$,0) point in reciprocal space and is observed to soften, and
broaden, away from this point. Samples with x=0, 0.01, 0.05, and
0.17 were studied. The new mode is found to be rapidly suppressed
with increasing Sr content and is absent at $x$=0.17, where it is
replaced by a continuum of excitations. The peak is only observed
when the incident x-ray polarization is normal to the CuO planes and
is also present in Nd$_2$CuO$_4$. We suggest possible explanations
for this excitation.

\end{abstract}


\maketitle

The high-frequency dynamics ($\hbar\omega \gtrsim J$) of the cuprate
superconductors are presently attracting intense interest for a
number of reasons. First, attempts to understand the dynamics within
the context of effective, spin-only Hamiltonians have raised several
outstanding questions, including understanding the importance of
higher-order corrections to the nearest-neighbor Heisenberg
Hamiltonian \cite{Sugai90,Coldea-PRL-01,Toader05,cyclic_comment05}.
Further, interest in the optical community has been focussed on the
same mid-IR energy regime, following claims that the
Ferrell-Glover-Tinkham sum rule is violated in several materials
\cite{Molegraaf02,Kuzmenko03,Boris04,Kuzmenko05}. While it is not
yet clear what the ultimate significance, if any, of these various
results will be in understanding the mechanism of high-T$_c$
superconductivity, it is clear that further experimental
investigations of this energy regime are vital. Here it will be of
particular value to investigate the momentum dependence of the
excitations to shed light on their character.

Motivated by this, we have carried out a resonant inelastic x-ray
scattering (RIXS) study of the mid-IR region as a function of
momentum transfer and doping in the La$_{2-x}$Sr$_x$CuO$_4$ system,
and the related cuprate Nd$_2$CuO$_4$. This technique, which probes
the electronic excitations associated with the copper site, provides
momentum-resolved, bulk property information and is therefore an
ideal probe for this endeavor. In the parent compound ($x$=0), we
find a previously unobserved excitation, peaked in {\bf q}-space
around the ($\pi$,0) point at 500 meV ($\sim$ 4000 cm$^{-1}$). On
moving away from the zone boundary, the excitation is observed to
soften and broaden. It is not observed at either the zone center
(0,0), or the antiferromagnetic point, ($\pi$,$\pi$). As a function
of Sr concentration, the excitation is rapidly suppressed,
broadening and weakening successively for x=0.01 and x=0.05. For
x=0.17, only the particle-hole continuum is observed. Further, we
find that the excitation has strong photon polarization dependence -
it is only observable when the incident polarization is along the
c-axis and not when it is in the $ab$ plane. Finally, investigations
of the related cuprate, Nd$_2$CuO$_4$, show a similar feature at 500
meV. Possible origins for this excitation are discussed.

The experiments were performed at beamline 9IDB, at the Advanced
Photon Source. A Si(111) monochromator and a Si(444) secondary
monochromator, provided an incident flux of $5 \times 10^{11}$ ph
s$^{-1}$, in a 0.1mm x 0.8mm (v x h) spot. The scattered radiation
was analyzed by a Ge(733) crystal. The overall resolution was 120
meV (FWHM) \cite{Hill07}. The samples were cooled to 20 K to reduce
the phonon contribution to the elastic scattering. Four samples of
La$_{2-x}$Sr$_x$CuO$_4$ (x=0, 0.01, 0.05 and 0.17) and $\rm
Nd_2CuO_4$ were studied. Except where noted, the incident
(horizontal) polarization was along the c-axis of the crystal.

Data were taken with the incident photon energy tuned to the peak of
the fluorescence following $1s \rightarrow 4p_{\pi}$ transitions,
where the $4p_{\pi}$ orbitals are perpendicular to the $\rm CuO_2$
plane \cite{energy-note}. Throughout this paper, we use the
tetragonal notation, that is a=b$\approx$ 3.79 \AA\ with {\bf
Q}=(100) parallel to the Cu-O-Cu bond direction. The x=0 and 0.17
samples had surface normals along the (100) direction, and the
x=0.01 and 0.05 samples had (110) surface normals. In all cases, we
report scattering in terms of the reduced wave-vector {\bf q}, where
{\bf Q=G+q}, with {\bf Q} the total momentum, and  {\bf G} a
reciprocal lattice vector \cite{Kim-cross_sect}.

\begin{figure}
\begin{center}
\includegraphics[angle=0,width=\figwidth]{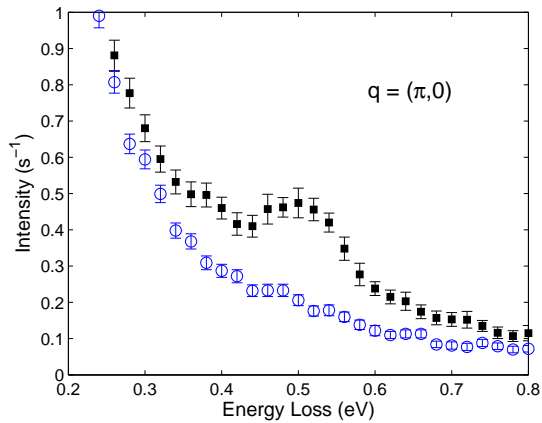}
\end{center}
\caption{(a) Inelastic x-ray scattering in La$_2$CuO$_4$ observed at
the ($\pi$,0) point, on and off resonance (closed and open symbols
respectively). A clear peak is observed at 500 meV on resonance.}
\label{fig1}
\end{figure}

\begin{figure}
\begin{center}
\includegraphics[angle=0,width=\figwidth]{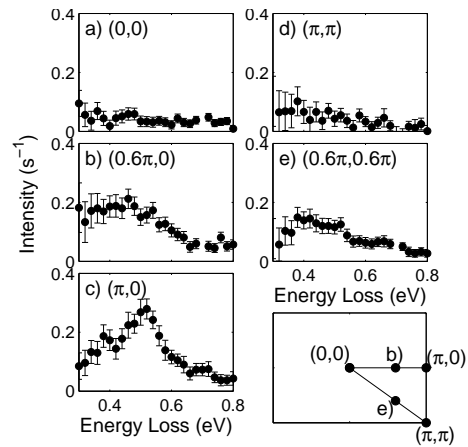}
\end{center}
\caption{Momentum dependence of the 500 meV feature in
La$_2$CuO$_4$. The figure shows the inelastic data, with the elastic
scattering subtracted off, for 5 different momenta in the 2D
Brillouin zone.} \label{fig2}
\end{figure}

In Fig.~\ref{fig1}, we show the resonant inelastic scattering in the
range 0.2 - 0.8 eV (closed symbols) at ($\pi$,0) in La$_2$CuO$_4$. A
clear peak of $\sim$ 0.2 counts per second is observed at 500 meV,
on a sloping background. The background arises from the tails of the
elastic scattering, which is 110 counts per second at zero energy
loss. In order to examine the peak in more detail, we subtract off
this elastic scattering - which arises from static disorder and a
small amount of thermal diffuse scattering. This is achieved by
tuning the incident photon energy to 8980 eV, far from resonance.
For all practical purposes, this ``turns off'' the inelastic
scattering and provides a clean measure of the elastic scattering.
We then subtract off this elastic scattering, appropriately scaled,
to obtain the resonantly enhanced inelastic signal. This is shown in
figure \ref{fig2}(c). A Lorentzian-like peak, of width 200 meV
(FWHM) is observed, peaked at 500 meV.

In Fig.~\ref{fig2}, we show the momentum dependence of this new
scattering in La$_2$CuO$_4$. Significantly, the new mode is not
observed at the zone center, (0,0), or  at the antiferromagnetic
point, $(\pi,\pi)$. Either the peak is absent at these points, or it
has dispersed to lower energies and is unobservable with the present
resolution. On moving away from $(\pi,0)$, the peak is seen to
soften slightly and broaden, with a peak position of 430 meV at
(0.6$\pi$,0) and (0.6$\pi$,0.6$\pi$) and a full width at these
points of 360 meV and 280 meV, respectively.

\begin{figure}
\begin{center}
\includegraphics[angle=0,width=\figwidth]{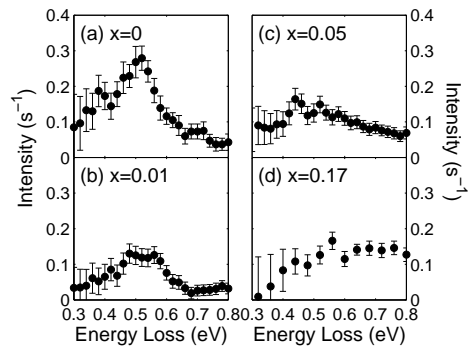}
\end{center}
\caption{Doping dependence of the 500 meV feature. Data are taken at
($\pi$,0) and have the elastic scattering subtracted off.}
\label{fig3}
\end{figure}

The doping dependence of the new feature is presented in
Fig.~\ref{fig3}, in which the scattering at $(\pi,0)$ is shown for
each of the four samples. The peak is observed to rapidly broaden
and weaken with increasing Sr content. The position does not change
to within errors. There is no measurable peak in the $x$=0.17
compound, where the particle-hole continuum due to metallic
electronic structure is observed. We note that for x=0.01 and
x=0.05, the same systematics in the momentum dependence of the peak
were observed as in the undoped compound, and preliminary data at
x=0.07 show similar behavior to x=0.05. The polarization dependence
of the scattering was also investigated. In Fig. 4, data are shown
for La$_2$CuO$_4$ with the incident photon polarization parallel to
c ($\epsilon \parallel \hat{c}$) and for $\epsilon \perp \hat{c}$.
The peak at 500 meV is only observed when the incident polarization
is along the c-axis, that is the photoelectron is in the Cu
4$p_{\pi}$ orbital in the intermediate state.

We now discuss possible origins for this new feature. First, it is
clear that it is not directly related to the mid-IR peaks seen in
optical conductivity measurements \cite{opt_ref}. This is because
the x-ray peak is not observed at the zone center, where the mid-IR
peaks are seen in the optical measurements, and because the strength
of this new feature increases as the Mott-Hubbard insulator state is
approached, in contrast to the doping dependence of the optical
measurements \cite{opt_ref}. These same systematics also rule out
the possibility that the peak arises from charge excitations
associated with any electronic inhomogeneity of the doped carriers
\cite{Homes03}. Finally, one could postulate that the peak results
from transitions to impurity-derived, mid-gap states. However, such
states would be expected to be highly localized and it is difficult
to explain the momentum-dependence, or the observed doping
dependence dependence. We conclude that this scattering is unlikely
to arise from such states. This leaves two plausible origins for
this new mode. One is that it is a {\em d-d} excitation, the other
that it is a multi-magnon process. We first consider {\em d-d}
excitations.

While most calculations put all {\em d-d} excitations at
significantly higher energies than discussed here (e.g. 1 eV
\cite{Graaf-PRB-2000}), it has been suggested that at least one is
much lower.  Perkins {\em et al.} explained a broadband mid-IR
absorption peak in terms of a low energy (0.4 eV) {\em d-d}
excitation together with phonon and magnon sidebands
\cite{Perkins1+2} and Ghiringhelli {\em et al.} recently calculated
the $d_{x^2-y^2} \rightarrow d_{3z^2-r^2}$ excitation to be 0.4 eV
in La$_2$CuO$_4$ \cite{Ghiringhelli-PRL-2004}. Such a low value is
controversial however, and most have argued that it is much higher,
for example Kuiper {\em et al.} suggested it was at 1.7 eV
\cite{Kuiper} --  though there has been no direct observation of
this excitation. If this present work is such an observation, then
one might explain the doping dependence observed here as the result
of an increased hybridization of the d-states as the metallic state
is approached \cite{Liu-PRL-1993}. The observed q-dependence is more
problematic for this picture, since one would expect such a local
excitation to show no dispersion, though it could arise from
anisotropic hopping.

To further explore this explanation for the 500 meV mode, we carried
out a study of Nd$_2$CuO$_4$, which lacks the apical oxygens of
La$_2$CuO$_4$ and has a 4\% larger in-plane CuO distance. One might
expect that this would shift the $d_{3z^2-r^2}$ orbitals to much
higher energies. In Fig. 4, we show the results for Nd$_2$CuO$_4$.
Just as in La$_2$CuO$_4$, a weak peak is observed at 500 meV. Thus
if this peak is the $d-d$ excitation, then these data require that
the changes in the oxygen octahedron cancel in the $d_{3z^2-r^2}$
energy. While this is perhaps unlikely, it is possible and thus we
cannot rule out this possible explanation of the new mode.

We now address the multi-magnon possibility. Two-magnon processes
have been studied in conventional Raman scattering experiments for
many years. In these experiments, a peak is observed at 375 meV
\cite{Lyons_PRB_88} and is associated with the creation of two
zone-boundary magnons with an energy reduced to $\sim 2.7$J by
magnon-magnon interactions
\cite{Elliot-PRL-1968,Chubukov95,Blumberg96,Lorenzana95,Freitas}.
Raman experiments, however, probe the {\bf q}=0 response (i.e the
total momentum of the magnon pair is zero) and so cannot be directly
compared to the present peak at ($\pi,0$). One can calculate the
momentum dependence of the non-interacting two-magnon
density-of-states (DOS) from the single magnon dispersion
\cite{Coldea-PRL-01}. This does show a peak at 500 meV at ($\pi$,0)
\cite{Coldea-private} and it is tempting to associate the present
x-ray peak with such scattering.  The DOS also has strong peaks at
600 meV at (0,0) and ($\pi,\pi$) \cite{Coldea-private}, which are
not seen in our data. This absence is may be due to the effect of
RIXS matrix element. In fact, recent calculations which suggest that
the two-magnon RIXS cross-section is zero or greatly reduced for
these points \cite{JvdB-private,vernay07,donkov07,nagao07}.

\begin{figure}
\begin{center}
\includegraphics[angle=0,width=2.85in]{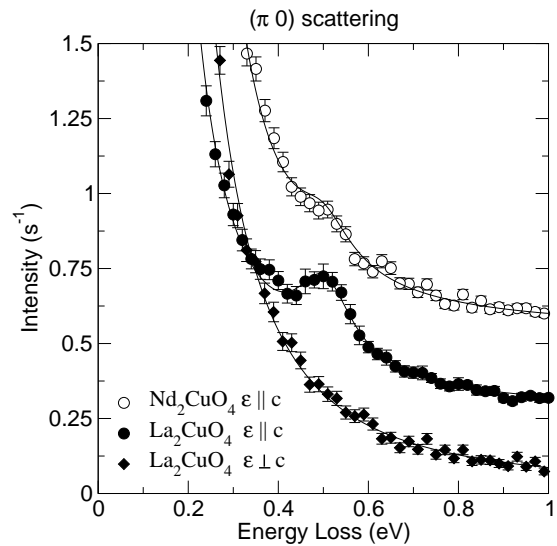}
\end{center}
\caption{Polarization dependence of the 500 meV peak in
La$_2$CuO$_4$ (closed symbols). The peak is only observed with the
incident polarization parallel to the c-axis. Data taken for
Nd$_2$CuO$_4$ (open circles) shows a weak peak at same energy. Scans
are shifted vertically for clarity.} \label{fig4}
\end{figure}

\begin{figure}
\begin{center}
\includegraphics[angle=0,width=\figwidth]{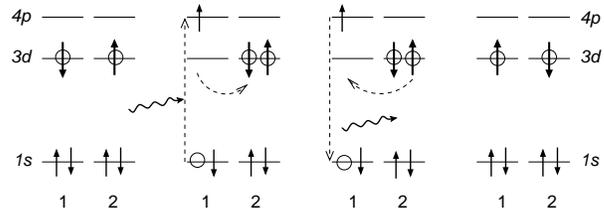}
\end{center}
\caption{Possible resonant scattering process resulting in the
creation of two magnons. The Cu (hole) spin on site 1 is repelled
onto a neighboring site, site 2, by the 1s core hole in the
intermediate state. Following the decay of the core-hole, the
``wrong'' spin hops back resulting in spin flips on both sites.
Note, the two sites could, in principle, be in neighboring layers.}
\label{fig5}
\end{figure}

If this new mode is indeed magnetic in origin, then it is natural to
ask how does RIXS couple to magnetic excitations. Figure~\ref{fig5}
shows one possible scattering mechanism. The system begins in the
ground state, with nearest neighbor 3d$^9$ spins
antiferromagnetically coupled: $|3d^9\uparrow ;3d^9\downarrow
>$. A $1s$ core-level electron is then excited into the $4p$ band.
The resonance utilized in these experiments is that of the
``well-screened'' intermediate state,
$|\underline{1s}3d^{10}{\underline L}4p>$, in which charge has moved
in to screen the core-hole from the oxygen ligand state. Further, it
is energetically favorable for this ligand hole to form a Zhang-Rice
singlet on the neighboring site \cite{Veenendaal}. In
Fig.~\ref{fig5}, we have represented this  as
$|\underline{1s}3d^{10}4p;3d^{8}\uparrow\downarrow >$. When the $4p$
decays, the ``wrong'' spin Cu hole hops back, and the net effect is
to flip two spins, leaving the system in a final state
$|3d^9\downarrow ;3d^9\uparrow>$. Other two-magnon processes are
also possible \cite{Platzman98}, though since these involve
quadrupole transitions, we deem them less likely.

The doping dependence of the excitation at ($\pi$,0) supports a
multi-magnon identification. In particular, the Raman data of Sugai
\cite{Sugai-PRB-2003} show a similar doping dependence for the two
magnon scattering in La$_{2-x}$Sr$_x$CuO$_4$ as observed here for
the x-ray peak, suggesting that both x-ray and Raman peaks may arise
from related processes.

One problem with the association of the 500 meV mode with two magnon
scattering is that the peak occurs at the energy expected for
non-interacting magnons.  While neutron scattering is known to be
insensitive to magnon-magnon interaction effects \cite{Huberman-05},
these are expected to be present in inelastic x-ray scattering,
since the two magnons are created in close proximity to one another
(See Fig.~\ref{fig5}). At finite momentum, Lorenzana and Sawatzky
have shown that such interactions can result in an almost bound
magnon-magnon resonance \cite{Lorenzana95}. This gives a sharp
quasi-particle-like peak at (${\pi,0}$). However, this peak does not
appear to explain the present {\bf q}-resolved data, since the
``bimagnon'' resonance at ($\pi$,0) occurs at $2.7J \sim 375$ meV. A
possible explanation in light of the incident polarization
dependence is that the peak results from two spin flips occurring in
adjacent $\rm CuO_2$ layers -- an {\em interlayer} bimagnon. Such
spin flips would only interact through small interlayer coupling and
would thus occur at essentially the non-interacting value.

Finally, we note that, motivated by the present experimental work, a
number of authors have investigated the magnetic inelastic x-ray
scattering cross-section. Notable amongst these is the work of
van~den~Brink who followed a perturbation expansion of the K-edge
RIXS spectra and finds quantitative agreement with the present
results \cite{JvdB-private}. Similarly, Nagao and Igarashi
\cite{nagao07} calculated two-magnon K-edge RIXS scattering and
others have investigated the finite-{\bf q} Raman cross-section
\cite{donkov07,vernay07}. All find qualitative similarities with the
symmetries of the present data.

In summary, we have observed a new collective mode in the mid-IR
response of La$_{2-x}$Sr$_x$CuO$_4$. Utilizing resonant inelastic
x-ray scattering, we show that the mode is peaked at the ($\pi$,0)
points, broadening and softening away from the zone boundary. The
mode is rapidly suppressed on doping with Sr. It is only present
with the incident polarization along the c-axis and is also present
in Nd$_2$CuO$_4$. We suggest that this mode is either the long
sought $d_{x^2-y^2} \rightarrow d_{3z^2-r^2}$ orbital excitation or
a two-magnon excitation, wtih perhaps the preponderance of evidence
in favor of a two-magnon process. While definitive identification of
this mode will have to wait further experimental and theoretical
investigations, the significance of this result lies in the
observation of this new mode in this important energy regime and the
potential for new insights on the high-energy Hamiltonaian that
understanding such a mode will provide.

\begin{acknowledgements}

We would like to thank J. van~den~Brink, A. Boothroyd, R. Coldea,
R.A. Cowley, J. Lorenzana, M. Turlakov and M. van Veenendaal for
invaluable discussions. The work at Brookhaven was supported by the
U. S. Department of Energy, Division of Materials Science, under
contract No. DE-AC02-98CH10886. The work at Toronto was supported by
NSERC. The work at U. Maryland was supported by NSF, DMR-0352735.
Use of the Advanced Photon Source was supported by the U. S.
Department of Energy, Basic Energy Sciences, Office of Science,
under contract No. W-31-109-Eng-38.

\end{acknowledgements}


\begin{thebibliography}{10}
\bibitem{energy-note}
The absolute energy is calibrated by measuring the absorption of a
pure Cu metal foil. In earlier experiments, the peak of the first
derivative for the metal spectra was taken to be at 8979 eV, which
put the resonance energy at 8991 eV. For the present work a more
recent value of 8980.5 eV was used, which puts the resonance energy
at 8992.5 eV. We stress that this is in fact the same energy as used
in earlier work and just represents a recalibration of the
spectrometer.
\end{thebibliography}
\end{document}